\documentclass[12pt]{iopart}
\usepackage{iopams}   
\usepackage[colorlinks=true,urlcolor=blue,citecolor=blue,linkcolor=blue]{hyperref}
\usepackage[T1]{fontenc}
\usepackage[latin9]{inputenc}
\usepackage{amssymb}
\usepackage{graphicx}
\usepackage{color}
\usepackage{mathrsfs}
\usepackage{float}
\usepackage{indentfirst}
\usepackage{color}

\usepackage{amsthm}
\usepackage{hyperref}
\usepackage{amscd}
\usepackage[english]{babel}

\usepackage{tikz}
\usetikzlibrary{arrows}

\newcommand{\ket}[1]{\left|{#1}\right\rangle}

\begin{document}
\title{Formation of Ultracold NaRb Feshbach Molecules}

\author{Fudong Wang, Xiaodong He, Xiaoke Li, Bing Zhu, Jun Chen and Dajun Wang}

\address{Department of Physics, the Chinese University of Hong Kong,\\
 Shatin, Hong Kong SAR, China}

\ead{djwang@phy.cuhk.edu.hk}

\begin{abstract}
\label{sec:Abstract}

We report the creation of ultracold bosonic $^{23}$Na$^{87}$Rb Feshbach molecules via magneto-association. By ramping the magnetic field across an interspecies Feshbach resonance, at least 4000 molecules can be produced out of the near degenerate ultracold mixture. Fast loss due to inelastic atom-molecule collisions is observed, which limits the pure molecule number, after residual atoms removal, to 1700. The pure molecule sample can live for 21.8(8) ms in the optical trap, long enough for future molecular spectroscopy studies toward coherently transferring to the singlet ro-vibrational ground state, where these molecules are stable against chemical reaction and have a permanent electric dipole moment of 3.3 Debye. We have also measured the Feshbach molecule's binding energy near the Feshbach resonance by the oscillating magnetic field method and found these molecules have a large closed-channel fraction.  
\end{abstract}

\maketitle

The creation and manipulation of ultracold heteronuclear molecules have received intensive attentions in recent years due to the versatile and promising potential applications~\cite{Carr2009,Krems2009} of these molecules. With controllable, anisotropic and long range dipole-dipole interactions, they could be used in quantum computation~\cite{DeMille2000,Andre2006}, quantum simulation~\cite{Micheli2006}, precision measurement~\cite{Zelevinsky2008,Hudson2011} and controlled cold chemistry ~\cite{Krems2008}. So far, the most successful scheme for producing ultracold ground-state dipolar molecule is by associating ultracold atoms near Feshbach resonances~\cite{Kohler2006,Chin2010} to form weakly-bound molecules first, followed by a stimulated Raman adiabatic rapid passage (STIRAP)~\cite{Bergmann1998} to transfer them to a deeply bound state~\cite{Winkler2007,Ospelkaus2008}. This has been successfully applied to the $^{40}$K$^{87}$Rb system~\cite{Ni2008}, where near degenerate ground-state dipolar fermionic molecules are created. However, the chemical reaction 2 KRb$\rightarrow$ K$_2$ + Rb$_2$ is an exoergic process which results in a large inelastic loss, severely limiting the trap lifetime of the KRb molecular gas~\cite{Ospelkaus2010,Ni2010,Miranda2011}.

Currently, there is a great effort in generalizing the KRb production scheme to other heteronuclear alkali dimers. Creation of Feshbach molecules of RbCs~\cite{Takekoshi2012}, LiNa~\cite{Heo2012} and NaK~\cite{Wu2012} were already reported in 2012, and very recently ground-state RbCs molecules were successfully produced~\cite{Takekoshi2014, Molony2014}. In this work, we focus on the bosonic NaRb molecule, which in the absolute ground state is stable against chemical reactions~\cite{Zuchowski2010} and has a permanent electric dipole moment as large as 3.3 Debye~\cite{Aymar2005}. The ground-state NaRb molecule can be readily polarized with a moderate electric field. For instance, at 5 kV/cm the induced dipole moment is already more than 2 Debye. Therefore, it is an appealing system for studying the bosonic quantum gas with strong dipolar interactions. Recently, the double species Bose-Einstein condensates (BECs) of $^{23}$Na and $^{87}$Rb atoms have been produced \cite{Xiong2013} and their interspecies Feshbach resonances(FRs) \cite{Wang2013} were also investigated in our group. One of the s-wave resonances between atoms in their lowest hyperfine Zeeman state locates conveniently at a magnetic field $B_0$ = 347.7 G with a width of $\Delta$ = 4.9 G. Here we report the creation and characterization of $^{23}$Na$^{87}$Rb Feshbach molecules by magneto-association with this FR.

We prepare the ultracold mixture of $^{23}$Na and $^{87}$Rb atoms in the same setup for producing the double species BEC \cite{Xiong2013}. Here we only describe it briefly. The initial evaporative cooling is performed in a hybrid trap \cite{Lin2009} with $^{87}$Rb as the coolant and the minority $^{23}$Na is sympathetically cooled. The sympathetic cooling is quite efficient thanks to the ideal heteronuclear scattering length of 66.8 $a_0$~\cite{Wang2013}, with $a_0$ the Bohr radius. The atoms are then transferred to a crossed optical dipole trap (ODT) where the final evaporative cooling is performed. In our previous work, the ODT was formed by a 1070 nm laser which generates very different trapping potential depths for Na and Rb, with a ratio {U$_\textrm{Na}$/U$_\textrm{Rb}$} $\approx$ 1/3. Thus when lowering the ODT power, Na always evaporates faster than Rb, limiting the double BEC numbers to about 4$\times 10^4$ for each species. Here, we add in another single beam trap created by a 660 nm laser (Cobolt Flamenco), which generates a repulsive potential for Rb and an attractive one for Na. In this dichromatic ODT, we can tune the relative trap depths almost at will to allow the majority Rb atoms to always act as the coolant in the ODT and minimize the loss of Na atoms. With this upgrade, the double BEC atom numbers are increased to more than 10$^5$ for both species.

For the best atom-to-molecule conversion, it is crucial for the two atomic clouds to have the maximum overlap in phase space~\cite{Hodby2005}, which ideally should be achieved with both species condensed. However, we have found empirically in our experiment that the largest pure Feshbach molecule sample can only be obtained with a near-degenerate mixture. This is believed to be a result of the very fast inelastic losses caused by collisions with the high density residual atoms inherent from the condensates. We also observe that better results can be obtained by turning off the 660 nm beam adiabatically at the end of the evaporative cooling. This is understood as a result of the smaller differential gravitational sag and thus better spatial overlap between the two clouds, as trap oscillation frequencies for the two species are much more similar in the pure 1070 nm ODT than in the dichromatic trap.

During evaporation, both atoms are transferred to their $|F = 1,m_{F} = 1\rangle$ hyperfine ground state with a radio-frequency adiabatic rapid passage in a 2 G magnetic field. The magnetic field is then ramped up to 364 G, far above the 347.7 G FR. Evaporation continues at this field and stops near the critical temperature $T_c$ of the Na BEC. We then jump the magnetic field down to 350 G, the starting point of the magneto-association. After 5 ms holding for the magnetic field to stabilize, we typically have an ultracold mixture with $8 \times 10^{4}$ Na and $1 \times 10^{5}$ Rb atoms at a temperature of $350$ nK. The average final trap oscillation frequency is measured to be $\overline{\omega} = 2\pi \times 170$(148) {Hz} for Na(Rb). The calculated peak number density and phase space density are $4 \times 10^{12}$($ 2.6 \times 10^{13}$) cm$^{-3}$ and 1.0(0.84) for Na(Rb), respectively.   

\begin{figure}[th]
\centering
\includegraphics[width=.7\linewidth]{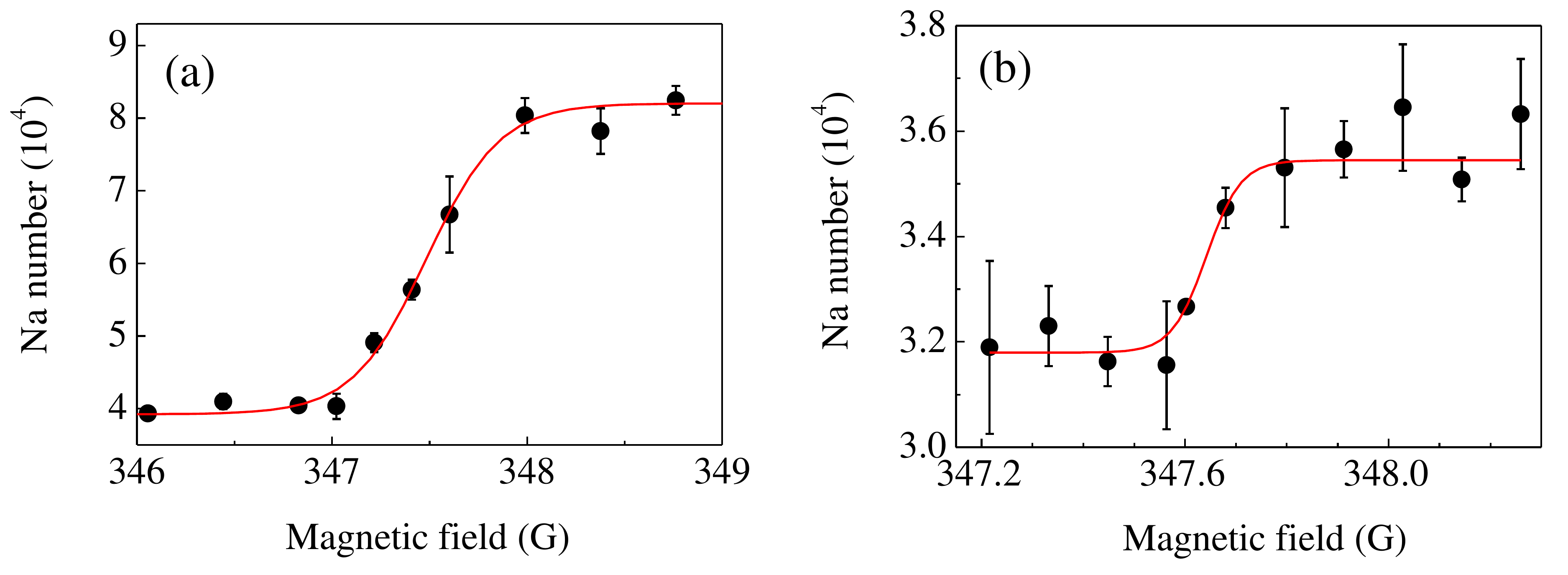}
 \caption{\label{fig:Association} Magneto-association and dissociation. (a) Remaining atoms after the magnetic field is swept downward and stopped at various values near the FR. (b) Reverse the sweeping direction without holding after the association, an atom recovery shows up as a result of molecule dissociation. Due to limited reverse ramp rate, we stop the association close to resonance at 347.2 G to shorten the dissociation sweep duration. Red solid lines are fit to the hyperbolic tangent function. Error bars are from statistics of typically 3 shots and represent one standard deviation.}
\end{figure}     

To perform magneto-association, we adiabatically sweep the magnetic field toward the FR and across it at a fixed rate of 5.2 G/ms. As shown in Fig.~\ref{fig:Association}(a), we stop at various field values and then turn the field off rapidly without holding before measuring the number of remaining atoms. Huge losses are observed for both species after the FR is crossed, amount to about 50$\%$ or $4 \times 10^{4}$ atoms for the minority Na atom. We note that these losses are the combined result of Feshbach molecule creation and enhanced three-body recombination. To confirm Feshbach molecules are indeed created, we ramp the field back up right after the association with a fixed rate of 3.9 G/ms. During this process, any Feshbach molecules will be dissociated to free atoms and show up as an atomic number recovery. A typical dissocaition curve is shown in Fig.~\ref{fig:Association}(b). An increase in the number of atoms is clearly observed after the field is ramped up across the FR, which verifys that Feshbach molecules were created and then dissociated. However, only 10$\%$ or 4000 of the lossed atoms were recovered. We believe that this is a lower bound of the Feshbach molecule number as fast atom-molecule collisions during the finite ramp up time can kill these molecules.


\begin{figure}
\centering
\includegraphics[width=.7\linewidth]{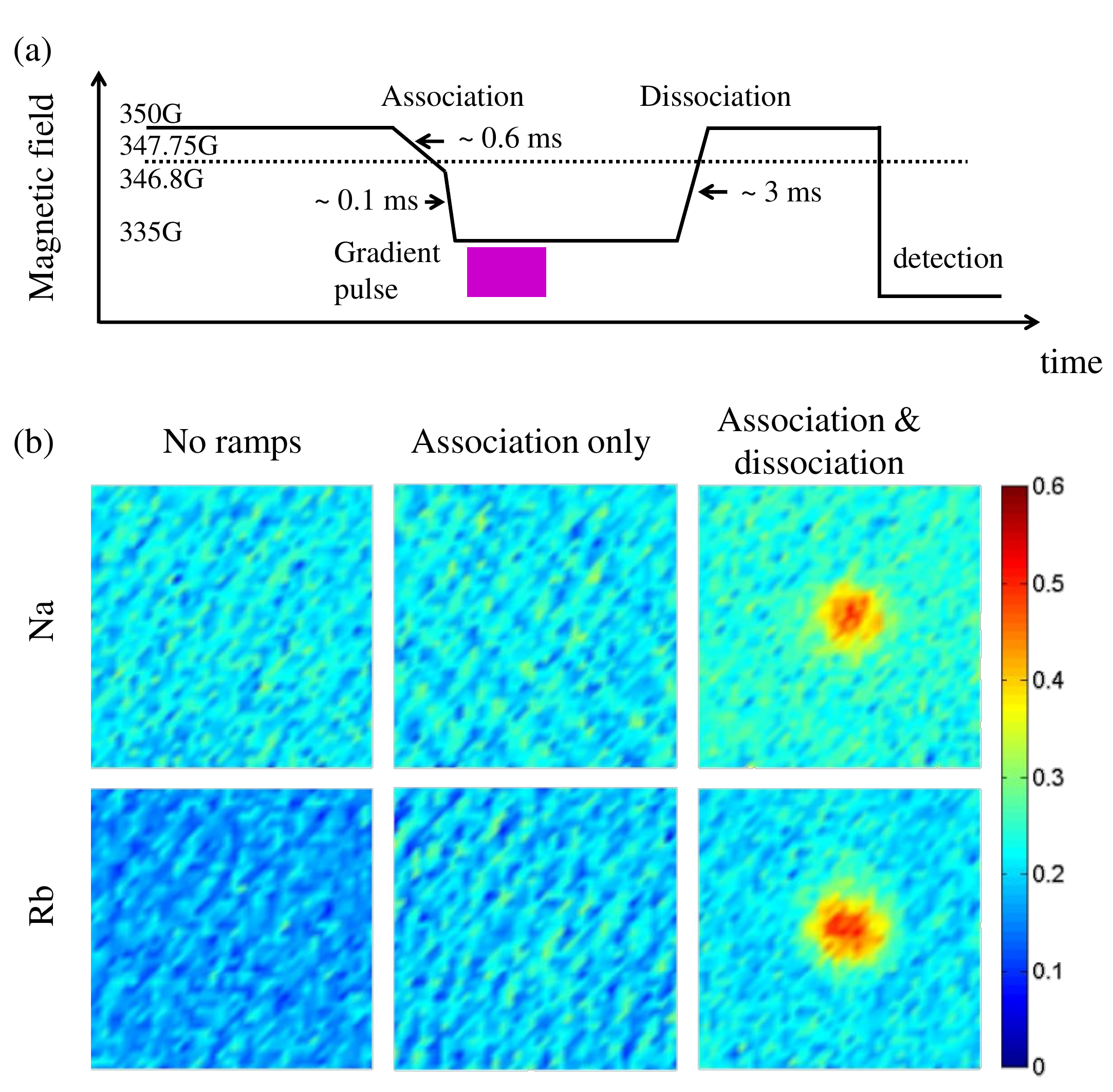}
 \caption{\label{fig:Image} 
 Creation of pure Feshbach molecule samples by removing residual atoms with a high magnetic field gradient after association. (a) shows the magnetic field sequence and (b) shows images of Na and Rb at different stages of the association and dissociation procedure with the removal gradient pulse always present(see text for details). Color bar denotes optical density. By varying the holding time before or after the gradient pulse, we can measure the lifetime of Feshbach molecules with or without residual atoms.}
\end{figure}

In future molecular spectroscopy studies and quest for absolute ground-state NaRb molecules, it is desirable to have a pure sample of Feshbach molecules. To this end, we have to remove residual Na and Rb atoms quickly after the association. According to the previous coupled-channel modeling \cite{Wang2013}, Feshbach molecules created with the 347.7 G FR have a nearly zero magnetic dipole moment when the binding energy is more than 11 MHz which corresponds to magnetic fields $\sim$7.5 G below resonance. As residual atoms in $\ket{1,1}$ state are high field seekers, we can thus remove them with a magnetic field gradient without affecting molecules. As illustrated in Fig.~\ref{fig:Image}, this is accomplished by a 3 ms gradient pulse of 150 G/cm with the Feshbach magnetic field at 335.0 G. After this, molecules are dissociated and the magnetic field is turned off before images are taken. With the high magnetic gradient applied, we have confirmed that both Na and Rb atoms can be removed completely without the association and dissociation ramps, or with the association ramp only, as evident by blank images in the first two columns of Fig.~\ref{fig:Image} (b). Signals appeared in the ``Association \& dissociation'' column of Fig.~\ref{fig:Image} (b) which are obtained after completing the full magnetic field sequence, can then only come from pure Feshbach molecules created by the association and survived the magnetic field gradient. In the optimized condition, we can produce a pure sample of 1700 NaRb Feshbach molecules which corresponds to a overall atom-to-molecule conversion efficiency (including contributions from both association and inelastic loss) of $2\%$.

\begin{figure}[th]
\centering
\includegraphics[width=.5\linewidth]{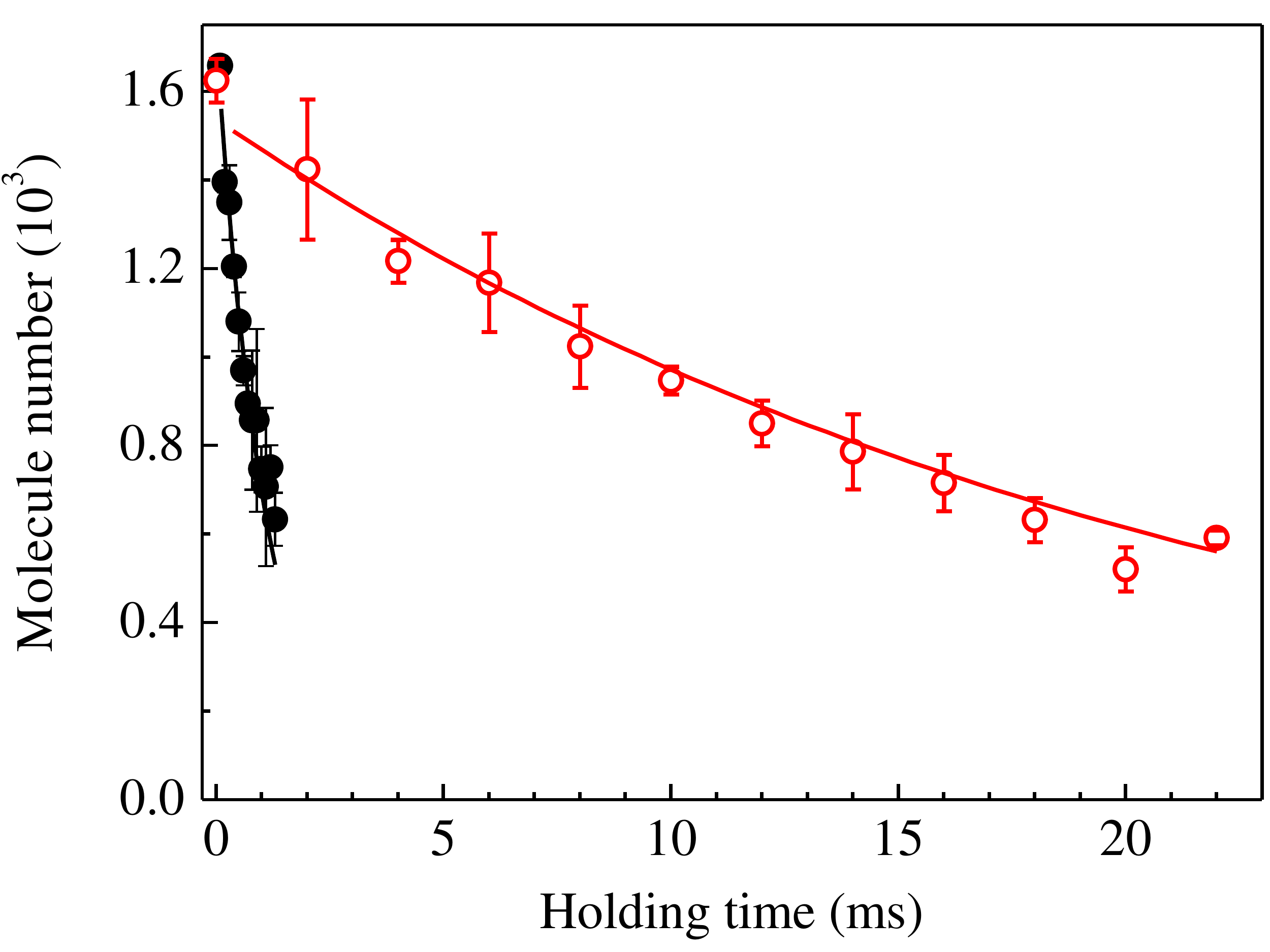}
 \caption{\label{fig:Lifetime} Lifetime of Feshbach molecules. Black solid circles are measured number of Feshbach molecules vs. holding time with residual Rb and Na atoms, while red open circles are without. Solid curves are exponential fitting for extracting lifetimes. Error bars represent one standard deviation from typically 3 shots. All measurements are performed at a magnetic field of 335 G, where Feshbach molecules are almost of pure closed-channel character (see discussions below).} 
\end{figure}

We notice that similar overall conversion efficiencies were also reported by other groups working on different bosonic heteronuclear molecular species \cite{Takekoshi2012, Koppinger2014}. Nevertheless, we want to emphasize that this low molecule number is not due to the association step only, which actually could have a much higher conversion efficiency. It is also a result of inelastic atom-molecule collisions. Even with the high field gradient, we estimate that about 1 ms is still needed for separating molecules from residual atoms completely, which is still ample time for inelastic collisions. As shown in Fig.~\ref{fig:Lifetime}, we have measured the Feshbach molecule lifetime with and without the residual atoms by varying the holding time before and after the gradient pulse. With residual atoms, the lifetime is 1.1(1) ms. In 1 ms, almost 60$\%$ of the population will be destroyed. With the residual atoms removed, the lifetime increases by a factor of twenty to 21.8(8) ms.

This lifetime and the current signal to noise ratio are good enough for further experiments towards ground state molecules. Faster residual atom removal with light blasting \cite{Thalhammer2006,Ni2008} should be able to increase the number of pure molecules. Ultimately, inelastic losses may be eliminated by first loading both atoms into three dimensional optical lattices and making a double species Mott insulator before the association \cite{Thalhammer2006,Ospelkaus2006,Chotia2012,Damski2003}. Currently, our numbers of atoms in the atomic mixture are also on the low side, but we cannot see any fundamental limitations which will prevent us from increasing them significantly by technical improvements.
  
\begin{figure}[th]
\centering
\includegraphics[width=.7\linewidth]{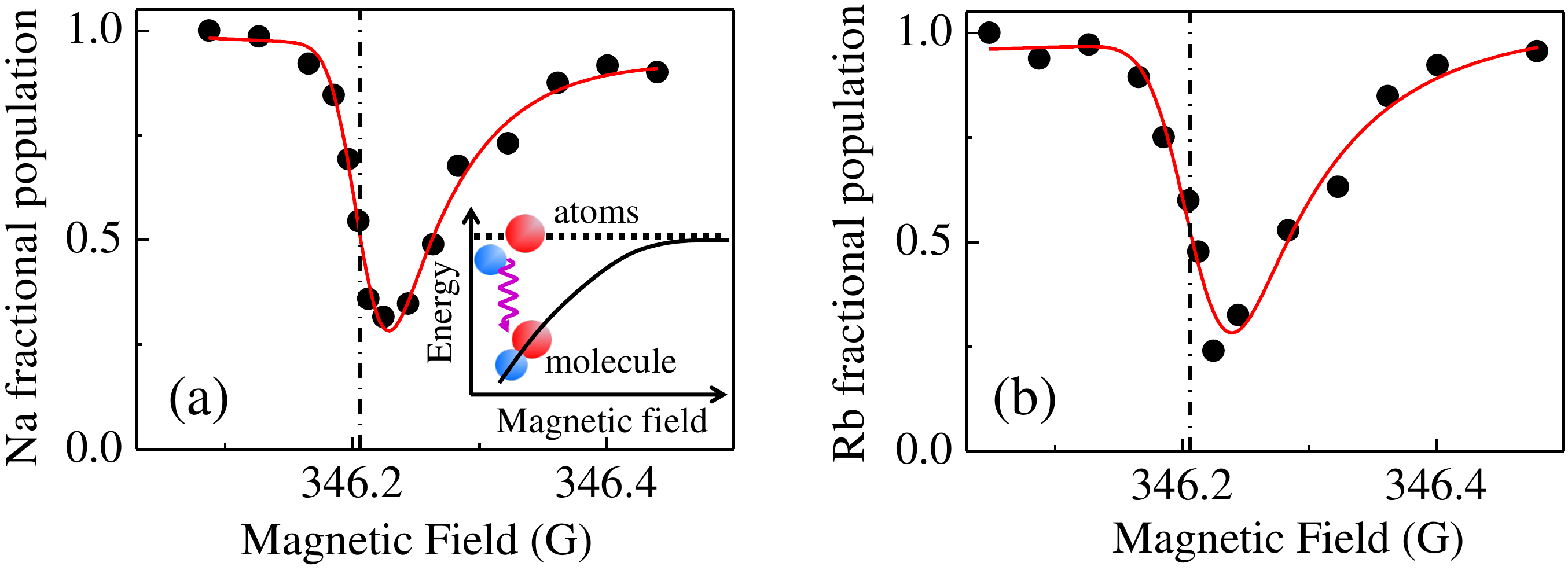}
 \caption{\label{fig:rfSpectroscopy} Feshbach molecule binding energy measurement by the oscillating $B$ field method. Inset of (a): an oscillating field can transfer atom pairs to a molecular state when its frequency matches with the Feshbach molecule's binding energy which is determined by the Feshbach field detuning. When scanning the Feshbach field in the presence of a 1 MHz oscillation field, besides the bare resonance at 347.7 G, additional loss features show up for both Rb (a) and Na (b) when the resonant condition is met. Solid dots are experimental data and solid curves are fitting results (see text). The dashed vertical lines indicate the resonance magnetic field found from the fit.}
\end{figure}

To further characterize the Feshbach molecule, we measure its binding energy vs. the magnetic field detuning with the oscillating field method \cite{Thompson2005}. The oscillating field is produced by a single loop coil driven by a 2 W radio frequency amplifier. It is placed coaxially with the Helmholtz coils producing the Feshbach field. As illustrated in the inset of Fig.~\ref{fig:rfSpectroscopy}(a), when the oscillation frequency matches with the Feshbach molecules' binding energy at the selected magnetic field detuning, transition between the free atom and Feshbach molecule states will be induced. This coupling can associate pairs of free atoms to Feshbach molecules, which will get lost from the trap quickly due to inelastic collisions.  

As the single loop coil has a rather narrow bandwidth, maintaining a constant field amplitude over all relevant frequencies is difficult. Similar to reference \cite{Takekoshi2012}, for the binding energy measurement, we instead fix frequency of the oscillating field, and scan the Feshbach magnetic field. As shown in Fig.~\ref{fig:rfSpectroscopy}(a) for Na and (b) for Rb are such a measurement in the presence of a 1 MHz oscillating field. Besides the main broad loss feature centering at 347.7 G, another narrower dip shows up on the lower field side when the Feshbach molecule's binding energy matches with the oscillation frequency. To extract the resonant magnetic field at this oscillating frequency, the apparently asymmetric lineshapes are fitted with a Gaussian function convolved with the Boltzmann distribution due to the finite temperature of our sample \cite{Hanna2007,Weber2008}. Typically, the center of the measured lineshape is shifted away from the resonance, as shown by the vertical lines in Fig.~\ref{fig:rfSpectroscopy}(a) and (b).  The same measurements are repeated for oscillation frequencies from 100 kHz to 3.5 MHz, beyond which the coupling becomes too weak for reliable measurement due to reduced free-bound Franck-Condon factors \cite{Chin2005}. 

\begin{figure}
\centering
\includegraphics[width=.7\linewidth]{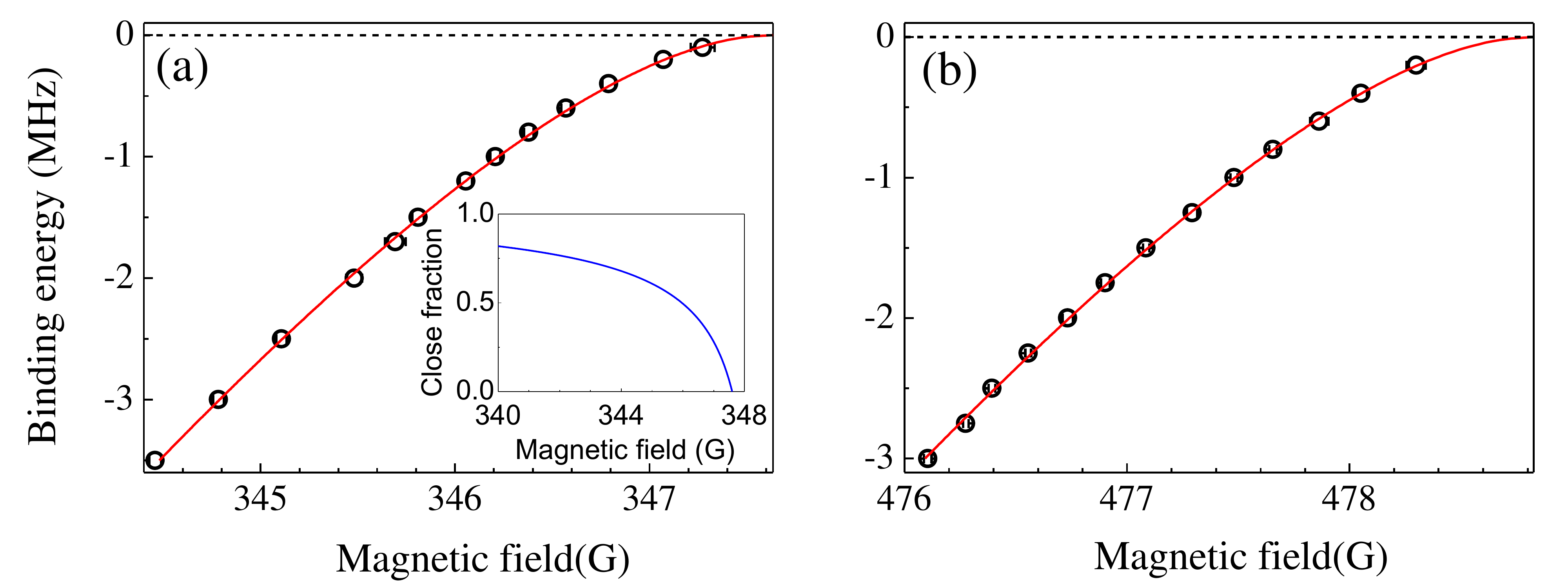}
 \caption{\label{fig:FigBinding} Binding energy vs. magnetic field near (a) 347.7 G and (b) 478.8 G FRs. Open circles are experimental data. Red solid lines are fitting results with the square well model and dashed lines indicate the thresholds. Error bars on magnetic field are from fitting and magnetic field calibration. The inset of (a) shows the closed-channel fraction of the Feshbach molecule vs. magnetic field near the 347.7 G FR calculated from the fitting parameters. }
\end{figure}

The results are summarized in Fig.~\ref{fig:FigBinding}(a). We have also performed the same characterization for another s-wave FR at 478.8 G \cite{Wang2013}, as shown in Fig.~\ref{fig:FigBinding}(b). We notice immediately that binding energies change quickly with magnetic field detuning for both resonances. Within 2 G, the Feshbach molecules are already bound by more than 3 MHz, which is a strong evidence of large closed-channel fraction or that these molecules are ``non-universal'' \cite{Kohler2006}. This finding is not surprising as the background scattering length is a positive and moderate value of 66.8 $a_0$, which indicates that closed channels for both FRs are real molecular states. We note that this behavior agrees qualitatively with the prediction from the coupled channel calculation \cite{Wang2013}. 

To extract quantitative information from the binding energy measurement, we fit our data with the square well model developed by Lange \textit{et al} \cite{Lange2009}. Near a Feshbach resonance with intermediate width, the binding energy $E_b=\frac{\hbar^2 k_m^2}{2\mu}$ can be determined from the magnetic dipole moment difference $\delta\mu$ between the open and closed channels and the Feshbach coupling strength $\Gamma$ as
\begin{equation}
k_m(B)=\frac{1}{a_{bg}-\overline{a}}+\frac{\Gamma/2}{\overline{a} \left(E_b+\delta\mu\left(B-B_c\right)\right)}.
\end{equation}
Here $k_m$ is the wave number, $B_c$ is the magnetic field where the bare molecular state is tunned to the open-channel threshold and $\overline{a}$ is the mean scattering length which can be calculated from the van der Waals $C_6$ coefficient. The solid curves in Fig.~\ref{fig:FigBinding} are the fitting results with $\delta\mu$, $\Gamma$, and $B_c$ as the fitting parameters. Here C$_6$ = 1.2946 $\times$ 10$^7$ cm$^{-1}$\r{A}$^{6}$ is used for calculating $\overline{a}$ and $a_{bg}$ is taken to be 66.77 $a_0$ \cite{Wang2013}. 

		\begin{table*}
		\caption{
		Parameters obtained from the square well model fitting of the binding energy vs. magnetic field data. Results from the coupled channel calculation in \cite{Wang2013} are also listed as a comparison. $\mu_B$ is the Bohr magneton.}
		\begin{center}
		\begin{tabular}{c c|c c c}
		\hline 
		\hline
		\multicolumn{2}{c|}{\textrm{coupled channel}}  & \multicolumn{3}{c}{\textrm{square well}}\\
		\hline 
		\textrm{B$_0$(G)} & \textrm{$\Delta$(G)} & \textrm{B$_0$(G)} & \textrm{$\Delta$(G)} & \textrm{$\delta\mu(\mu_B)$}\\
		\hline
		 347.75           & 4.89              & 347.64(3)      & 5.20(0.27)        &  2.66(29)        \\
		 478.79           & 3.80              & 478.83(3)      & 4.81(0.27)        &  2.52(26)        \\
		\hline \hline  
		\end{tabular}
		\label{table1}
		\end{center}
		\end{table*}

The resonance width $\Delta=\frac{1}{\delta\mu}\frac{\left(a_{bg}-\overline{a}\right)^2}{a_{bg}\overline{a}}\frac{\Gamma}{2}$, and resonance magnetic field $B_0=B_c-\frac{a_{bg}}{a_{bg}-\overline{a}}\Delta$ can then be obtained from combination of these parameters, as summarized in Table.~\ref{table1}. While $B_0$ for both resonances agree well with results from the global coupled channel modeling, the resonance widths show some discrepancy. We also want to mention that binding energies predicted from coupled channel calculations also have some disagreement with our measurement. We hope to resolve these problems by adding more data points to the Feshbach spectroscopy with improved resolutions later.

From the fitting parameters, the widely used dimensionless resonance strength parameter s$_{res}$ \cite{Chin2005} for the two FRs are calculated to be 0.72 and 0.63, respectively, which confirms that their coupling strengths are both in the intermediate regime. We have also deduced the closed channel fraction $\frac{1}{\delta\mu}\frac{\partial{E_B}}{\partial{B}}$ of the Feshbach state. As shown in the inset of Fig.~\ref{fig:FigBinding}(a), for the 347.64 G FR, the closed-channel character increases rapidly with the magnetic field detuning. At about 346 G, the closed-channel fraction is already 50$\%$ and reaches over 80$\%$ at 340 G.

In conclusion, we have successfully produced ultracold $^{23}$Na$^{87}$Rb Feshbach molecules by Feshbach magneto-association. After removing residual atoms, the small pure molecular sample lives long enough for further work. These Feshbach molecules have a large closed-channel fraction near the Feshbach resonance, advantageous for finding high efficiency STIRAP routes for population transfer. We note that a promising path for population transfer via the $2^1\Sigma^+/1^3\Pi$ excited state admixture was already investigated in detail with conventional molecular spectroscopy\cite{Docenko2007}. We are now in a good starting point toward chemically stable ground-state bosonic molecules with a large electric dipole moment of 3.3 Debye for investigating ultracold gas with dominating dipolar interactions. 

\section*{Acknowledgments} 

We thank E. Tiemann, O. Dulieu and G. Qu\'{e}m\'{e}ner for valuable discussions and Mingyang Guo for technical assistance. This work is supported by Hong Kong Research Grants Council (General Research Fund Projects 403111, 404712 and the ANR/RGC Joint Research Scheme ACUHK403/13).

\end{document}